\documentclass[3p,preprint]{elsarticle}

\usepackage{amssymb, amsfonts, latexsym, mathtools} %
\usepackage{graphicx, setspace, subfigure}
\usepackage{amsmath,euscript,mathrsfs}
\usepackage[hidelinks]{hyperref} 
\usepackage[usenames,dvipsnames,svgnames]{xcolor}

\hypersetup{
    colorlinks,
    linkcolor={blue!50!black},
    citecolor={blue!50!black},
    urlcolor={blue!50!black}
}

\graphicspath{{Figures/}}

\makeatletter
\def\blfootnote{\gdef\@thefnmark{}\@footnotetext}
\makeatother

\usepackage[all]{hypcap}

\usepackage{array}
\newcolumntype{C}[1]{>{\centering\arraybackslash}m{#1}}

\usepackage[super]{nth}

\usepackage{comment}

\allowdisplaybreaks[1]

\newcommand{\lb}{\left (}
\newcommand{\rb}{\right )}

\renewcommand{\O}[1]{O \lb #1 \rb}

\journal{Wave Motion}

\begin{document}

\begin{frontmatter}

\title{On Boussinesq-type models for long longitudinal waves in elastic rods} %
\author[spb]{F.E.  Garbuzov}
\author[lbo]{K.R. Khusnutdinova\corref{cor1}} 
\ead{K.Khusnutdinova@lboro.ac.uk}
\author[spb]{I.V. Semenova}
\cortext[cor1]{Corresponding author.}
\address[spb]{Ioffe Institute, 26 Polytekhnicheskaya, St. Petersburg 194021, Russia}
\address[lbo]{Department of Mathematical Sciences, Loughborough University, Loughborough LE11 3TU, United Kingdom\\[2ex]
\hspace{10cm} \bf In memory of Alexander M. Samsonov}


\begin{abstract}%
In this paper we revisit the derivations of model equations describing long nonlinear longitudinal bulk strain waves in elastic rods within the scope of the Murnaghan model in order to derive a Boussinesq-type model, and extend these derivations to include axially symmetric loading on the lateral boundary surface, and longitudinal pre-stretch.  We systematically derive two forced Boussinesq-type  models from the full equations of motion and non-zero surface boundary conditions, utilising the presence of two small parameters characterising the smallness of the wave amplitude and the long wavelength compared to the radius of the waveguide. We compare the basic dynamical properties of both models (linear dispersion curves and solitary wave solutions). We also briefly describe the laboratory experiments on generation of bulk strain solitary waves in the Ioffe Institute,  and suggest that this generation process can be modelled using the derived equations.
\end{abstract}


\end{frontmatter}

\section{Introduction}

The study of nonlinear waves in solids is an important theme of the current research on waves (see, for example, \cite{Maugin,Dai,M,HL,E1,P,E2} and references therein). The research includes the studies of bulk strain solitons in solid waveguides (e.g. \cite{S_book,P_book}). Historically, theoretical developments began with the studies of waves in elastic rods of circular cross section. G.A. Nariboli and A. Sedov have systematically derived the Burgers - Korteweg de Vries equation for long longitudinal waves in a viscoelastic rod using power series expansions in the radius \cite{NS}. Later, L.A. Ostrovsky and A.M. Sutin have developed a regularised Boussinesq - type model using the plane cross section and Love's hypothesis in order to simplify the Lagrangian of the problem \cite{OS}. 
A.M. Samsonov has suggested a model with two types of dispersive terms  \cite{S1}, and a model for the rod with a variable radius and elastic moduli \cite{S2}. The coefficients of Samsonov's model with two types of dispersive terms have been refined in the works by A.M. Samsonov and A.V. Porubov  \cite{SP, S_book, P_book}, and a dispersive-dissipative model has been suggested by A.V. Porubov and M.G. Velarde \cite{PV}. A model with three types of dispersive terms has been discussed by V.I. Erofeev et al. (see \cite{E_book} and references therein), however, the nonlinearity coefficient of this model differs from that in the  models of L.A. Ostrovsky and A.M. Sutin, and A.M. Samsonov and A.V. Porubov, while the choice of dispersive coefficients has not been fixed.

All derivations of Boussinesq-type models in these studies were based on the use of the Murnaghan model for elastic energy \cite{Murnaghan}, accounting for both physical and geometrical sources of nonlinearity, and subsequent simplification of the full Lagrangian of the problem using some hypothesis. In \cite{DF} H.-H. Dai and X. Fan have obtained a system of two coupled equations and a uni-directional model of the Benjamin-Bona-Mahony (BBM) type {\cite{BBM} within the scope of nonlinear elasticity, using a systematic asymptotic derivation from the full equations of motion and free surface boundary conditions. The systematic asymptotic analysis has been also developed for the description of linear transient waves in a pre-stretched compressible hyperelastic material (with application to Mooney-Rivlin material) in \cite{DC}. In  \cite{KSZ}, a Boussinesq-type model 
for a nonlinearly elastic waveguide with both physical and geometrical sources of nonlinearity 
has been derived using  the systematic asymptotic derivation within the scope of a lattice model, i.e. the model equation has been derived from the full equations of motion, and no hypothesis have been used to simplify the Lagrangian of the problem. Interestingly, Boussinesq-type models with both two and three types of dispersive terms have been  derived in the latter study, as well as coupled Boussinesq-type equations for the waves in a layered waveguide with an imperfect interface.

The Boussinesq-type models have been recently used to study the scattering of long longitudinal bulk strain solitary waves by delamination in \cite{KS,KT1, KT2}, and some related experimental observations have been reported in \cite{JAP2010, JAP2012}. We also note the related studies on nonlinear wave scattering by defects in lattices \cite{ML} and strings \cite{AM}. Some other Boussinesq-type models have been derived to describe the propagation of large amplitude transverse waves (see \cite{DS} and references therein).

The aim of our current paper is to revisit the derivation of nonlinear two-directional long wave models  for longitudinal waves within the scope of nonlinear dynamic elasticity in order to systematically derive a Boussineq-type model, 
and  to extend these derivations to the more complicated case when there is non-zero axisymmetric loading on the boundary surface, and a background longitudinal pre-stretch.
The derivations are performed using symbolic computations with MATHEMATICA \cite{Mathematica}. We also discuss the basic solitary wave solutions  and dispersive properties of the models.
Finally, we briefly describe the experiments on generation of solitary waves in the Ioffe Institute by the group which was headed by Alexander M. Samsonov, and suggest that our derived models will be useful for the mathematical modelling of the generation processes. This paper is dedicated to the memory of our dear colleague and friend.

\section{Problem formulation}

We consider a rod of circular cross section with the radius $R$ and use cylindrical coordinates $(x, r, \varphi)$ with the axial coordinate $x$, radial coordinate $r$ and angular coordinate $\varphi$. We use the Lagrangian description and denote the displacement vector by $\underline{U} = (U, V, W) $, where $ U $ is the axial displacement, $ V $ is the radial displacement and $ W $ is the torsion. 
\begin{figure}[h]
	\centering
	\includegraphics[width=0.35\textwidth]{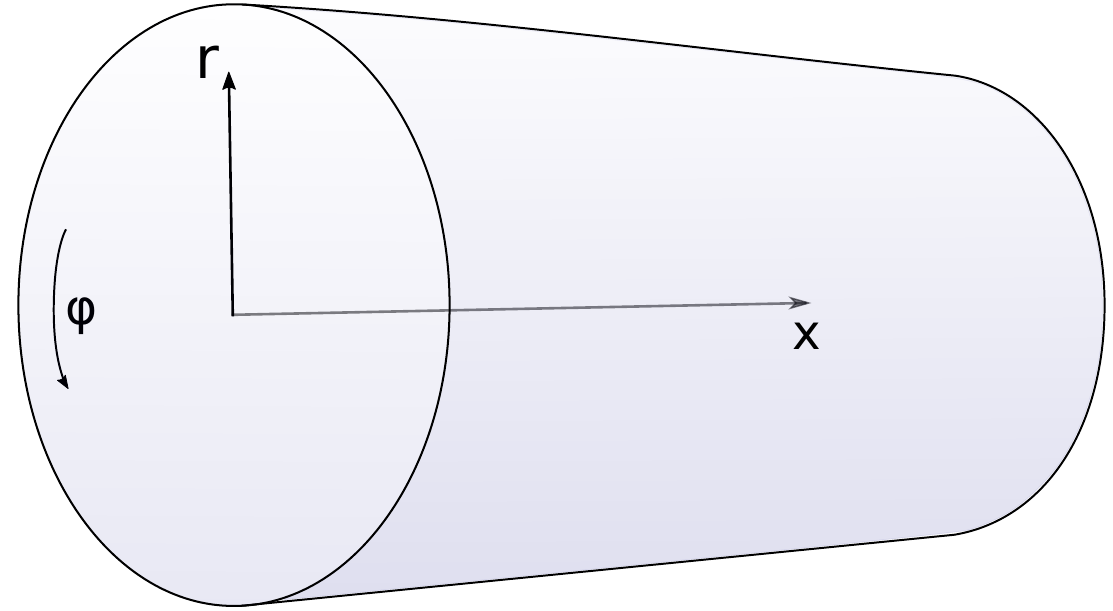}
	\caption{Schematic of the problem.}
	\label{fig:rod}
\end{figure}

Similarly to the previous studies discussed in the Introduction we assume that the rod is made of a Murnaghan's hyperelastic material (\cite{Murnaghan}, see also \cite{S_book} and references therein). The strain energy density of that material can be written as
\begin{equation}\label{murnaghan_pot}
\Pi = \frac{\lambda + 2\mu}{2}I_1(\underline{\underline{C}})^2 - 2\mu I_2(\underline{\underline{C}}) + \frac{l+2m}{3}I_1(\underline{\underline{C}})^3 - 2m I_1(\underline{\underline{C}}) I_2(\underline{\underline{C}}) + n I_3(\underline{\underline{C}}),
\end{equation}
where $I_1(\underline{\underline{C}}) = {\rm tr}\ \underline{\underline{C}},\  I_2(\underline{\underline{C}}) = [({\rm tr}\ \underline{\underline{C}})^2 - {\rm tr}\ \underline{\underline{C}}^2]/2,\ I_3(\underline{\underline{C}}) = {\rm det}\ \underline{\underline{C}}$ are invariants of the Cauchy-Green strain tensor 
\begin{equation}
\underline{\underline{C}} = \frac12( (\nabla\underline{U})^T + \nabla\underline{U} + (\nabla\underline{U})^T\cdot\nabla\underline{U}), 
\label{C}
\end{equation}
$\lambda$, $\mu$ are the Lame coefficients and $l$, $m$, $n$ are the Murnaghan moduli. Here, and in the following, all spatial derivatives are taken with respect to the natural (underformed) coordinates  in the reference configuration. It is worth noting that the Murnaghan model is used here in the sense of the general weakly-nonlinear theory of elasticity, assuming small values of strain (see the discussion in \cite{DSS}).

The equations of motion are given by
\begin{equation}
\rho\  \ddot {\underline{U}} = {\rm div}\ \underline{\underline{P}},
\label{P}
\end{equation}
where $\rho$ is the material density in the natural configuration, dots denote differentiation with respect to time, and $ \underline{\underline{P}} $ is the first Piola-Kirchhoff stress tensor 
\begin{equation}\label{piola-k}
\underline{\underline{P}} =  (I + \nabla\underline{U}) \cdot \frac{\partial\Pi}{\partial\underline{\underline{C}}}.
\end{equation}

We now consider an exact reduction of the full equations of motion describing solutions with no torsion ($W=0$), and where the longitudinal and transverse displacements $U$ and $V$ are independent of $ \varphi $:
\begin{equation}\label{assumptions}
W = 0, \quad U = U(x,r,t), \quad V = V(x,r,t).
\end{equation}

The equations of motion take the form
\begin{eqnarray}
\label{eq1_0}
\rho  \frac{\partial ^2 U(x,r,t)}{\partial t^2}-\frac{\partial P_{x x}}{\partial x}-\frac{\partial P_{xr}}{\partial r}-\frac{P_{xr}}{r} &=& 0,\\
\label{eq2_0}
\rho  \frac{\partial ^2 V(x,r,t)}{\partial t^2} - \frac{\partial P_{rx}}{\partial x}-\frac{\partial P_{rr}}{\partial r}-\frac{P_{rr} - P_{\varphi\varphi}}{r} &=& 0,
\end{eqnarray}
while the third equation is identically satisfied. Here, $P_{\alpha \beta}$ denote components of the first Piola-Kirchhoff stress tensor, and we do not use different fonts to explicitly indicate the difference between the current and reference configurations in the subscripts.

The equations of motion are complemented by the surface boundary conditions
\begin{equation}
\underline{\underline{P}}\cdot \underline{n} = \underline{P_b} ,
\label{bc}
\end{equation}
where $\underline{n}$ is the normal vector to the lateral surface, and $\underline{P_b} = (P(x,t), T(x,t), 0)$, i.e. we consider the rotationally symmetric case, and $P(x,t)$ and $T(x,t)$ are the known tractions.
Thus, the boundary conditions are given by
\begin{eqnarray}
	P_{rr} &=& P(x, t) \quad \mbox{at} \quad r = R \label{bc_rr},\\
	P_{xr} &=& T(x, t) \quad \mbox{at} \quad r = R \label{bc_rx}.
\end{eqnarray}
Since the component $ P_{\varphi r} \equiv 0 $, the third boundary condition $P_{\varphi r} = 0$ at $r = R$ is identically satisfied.

\section{Derivation of two forced Boussinesq-type equations} 
\label{s:power_series_exp}
Our approach  to the derivation of the model equation in this section is essentially similar to the derivation in \cite{DF}, but we simplify it by adopting expansions used in the derivation of the linear model in \cite{bostrm2000}.  Thus, we look for a solution of the  problem in the form of power series expansions of the displacements in the radial coordinate:
\begin{eqnarray}
\label{u_series}
U(x,r,t) &=& U_0(x,t) + r^2 U_2(x,t) + r^4 U_4(x,t) + \dots \, ,\\
\label{v_series}
V(x,r,t) &=& r V_1(x,t) + r^3 V_3(x,t) + r^5 V_5(x,t) + \dots \, .
\end{eqnarray}
Note that the longitudinal displacement is a series in even powers of the radius, while the transverse displacement is expanded in odd powers (see  \cite{bostrm2000}). Unlike \cite{DF}, in this paper we aim, firstly, to systematically derive a Boussinesq-type equation, and, secondly, to account for non-zero loading on the lateral surface of the rod, as well as uni-axial (longitudinal) pre-stretch.

We consider the waves of small amplitude and large length compared to the radius of the rod. Hence we non-dimensionalise the variables and tractions as follows:
\begin{equation} \label{scales1}
\tilde t = \frac{t}{L/c}, \quad \tilde x = \frac{x}{L}, \quad \tilde r = \frac{r}{\delta L}, \quad \tilde U = \frac{U}{\varepsilon L}, \quad \tilde V = \frac{V}{\varepsilon \delta L}, \quad \tilde P = \frac{P}{E \varepsilon}, \quad \tilde T = \frac{T}{E \varepsilon\delta},
\end{equation}
which yields
$\displaystyle \tilde U_n =  \frac{L^n U_n}{\varepsilon L},  \ \tilde V_n =  \frac{L^n V_n}{\varepsilon L}$ for $n \ge 0$, 
assuming that $L$ is the characteristic wavelength, $c$ is the linear wave speed, $E$ is the Young modulus, $\varepsilon$ is the small amplitude parameter (characterising the longitudinal strain), and $\displaystyle \delta = \frac{R}{L} $ is the second small parameter (long wavelength parameter). Here, the tilde denotes dimensionless variables and tractions. We are also interested in the equation for the case of weak tractions, which can be obtained by rescaling $\tilde P$ and $\tilde T$ in the derived equation.  In the following we will use the expressions for the Young modulus and the Poisson ratio in terms of the Lame coefficients:
\begin{equation}\label{young_mod}
E = \frac{\mu(3\lambda + 2\mu)}{\lambda + \mu}, \quad \nu = \frac{\lambda}{2 (\lambda + \mu)}.
\end{equation}

Then, the expansions \eqref{u_series} and \eqref{v_series} take the form
\begin{eqnarray}
\label{u_series_scaled}
U(x,r,t) &=& \varepsilon L \left(\widetilde{U}_0(\tilde x, \tilde t) + \delta^2 \tilde{r}^2\widetilde{U}_2(\tilde x, \tilde t) + \delta^4\tilde{r}^4\widetilde{U}_4(\tilde x, \tilde t) + O(\delta^6)\right),\\
\label{v_series_scaled}
V(x,r,t) &=& \varepsilon L \delta \left(\tilde{r} \widetilde{V}_1(\tilde x, \tilde t) + \delta^2 \tilde{r}^3\widetilde{V}_3(\tilde x, \tilde t) + \delta^4 \tilde{r}^5\widetilde{V_5}(\tilde x, \tilde t) + O(\delta^6)\right).
\end{eqnarray}
In what follows we omit the tildes.

Substituting \eqref{u_series_scaled} and \eqref{v_series_scaled} into the equations of motion \eqref{eq1_0}, \eqref{eq2_0} we obtain
\begin{equation} \label{eq1_1}
\begin{split}
&\rho c^2 U_{0tt} - (\lambda + 2\mu) U_{0xx} - 2(\lambda + \mu) V_{1x} - 4\mu U_2 + \Phi_1(U_0, V_1, U_2) \varepsilon\\
&+ \left[\rho c^2 U_{2tt} - (\lambda + 2\mu)U_{2xx} - 4(\lambda + \mu)V_{3x} - 16\mu U_4\right] \delta^2 r^2 + O(\varepsilon^2, \varepsilon\delta^2, \delta^4) = 0,
\end{split}
\end{equation}
\begin{equation} \label{eq2_1}
\begin{split}
&\left[\rho c^2 V_{1tt} - \mu V_{1xx} - 2(\lambda + \mu)U_{2x} - 8(\lambda + 2\mu)V_3 - \Phi_2(U_0, V_1, U_2, V_3)\varepsilon \right] r\\
&- \left[\rho c^2 V_{3tt} - \mu V_{3xx} - 4(\lambda + \mu)U_{4x} - 24(\lambda + 2\mu)V_5 \right] \delta^2 r^3  + O(\varepsilon^2, \varepsilon\delta^2, \delta^4) = 0.
\end{split}
\end{equation}
Here, the subscripts $x$, $y$ and $z$ denote partial derivatives. Nonlinear terms are given by 
\begin{equation*}
\begin{split}
&\begin{split}
\Phi_1 =& \, 2\left[(-4\lambda - 4\mu + n - 4m) V_1 - 2(\lambda + 2\mu + m) U_{0x}\right] U_2 - \left[ 2(2l + \lambda) V_1 + (3\lambda + 6\mu + 2l + 4m) U_{0x} \right] U_{0xx} \\
& - \left[ (2\lambda + 2\mu + 8l + n) V_1 + 2(\lambda + \mu + 2l + m) U_{0x} \right] V_{1x},
\end{split} \\
&\begin{split}
\Phi_2 =& \, \frac12 \left[2 (2\lambda + 2\mu + 8l + n) U_{2x} + (4\lambda + 4\mu + 4m - n) V_{1xx} + 32(2\lambda + 3\mu + 2l + 2m) V_3 \right] V_1 \\
& + 2(\lambda + \mu + 2l + m) U_{0x} U_{2x} + 2(\mu + m)\left[ U_{0xx} + 4 V_{1x} \right] U_2 + (\lambda + 2\mu + m)(U_{0x} V_{1x})_x\\
& + \frac14(12 \lambda + 20 \mu + 12m - n) V_{1x}^2 + 8(\lambda + 2l) V_3 U_{0x} + (4\lambda + 12 \mu + 4m + n) U_2^2.
\end{split}
\end{split}
\end{equation*}
The functions $ U_2 $, $ V_3 $, $ U_4 $ can be obtained from \eqref{eq1_1} and \eqref{eq2_1} by equating to zero the coefficients at different powers of $\delta$, and they have the following form:
\begin{eqnarray}
\label{U2}
U_2 &=& \frac{1}{4\mu} \left[ \rho c^2 U_{0tt} - (\lambda + 2\mu) U_{0xx} - 2(\lambda + \mu) V_{1x} \right] + \varepsilon f_2(x,t) + O(\varepsilon^2),\\
\label{V3}
V_3 &=& \frac{1}{8(\lambda + 2\mu)} \left[ \rho c^2 V_{1tt} - 2(\lambda + \mu) U_{2x} - \mu V_{1xx} \right] + \varepsilon f_3(x,t) + O(\varepsilon^2),\\
\label{U4}
U_4 &=& \frac{1}{16\mu}\left[\rho c^2 U_{2tt} - (\lambda + 2\mu) U_{2xx} - 4(\lambda + \mu) V_{3x}\right] + O(\varepsilon).
\end{eqnarray}
The expressions for the functions $f_2$ and $f_3$ are rather lengthy, and therefore they are not shown here.

Next, substituting the functions $ U_2 $, $ V_3 $, $ U_4 $ into the boundary conditions \eqref{bc_rr}, \eqref{bc_rx} we obtain the equations
\begin{equation} \label{bc_rr_subst}
\begin{split}
2 (\lambda + \mu) V_1 + \lambda U_{0x} + \varepsilon \Psi_1(U_0, V_1) 
& + \frac{\delta^2}{8} \bigg[ (\lambda + 3\mu) U_{0xxx}- \frac{\rho c^2(\lambda + 3\mu)}{\lambda + 2\mu} U_{0xtt} \\
&+ \frac{2\rho c^2(2\lambda + 3\mu)}{\lambda + 2\mu} V_{1tt} + 2\lambda V_{1xx}\bigg] 
+ O(\varepsilon^2, \varepsilon\delta^2, \delta^4) =  \frac{\mu(3\lambda + 2\mu)}{\lambda + \mu} P,
\end{split}
\end{equation}
\begin{equation} \label{bc_rx_subst}
\begin{split}
\rho  c^2 U_{0tt} -2 \lambda  V_{1x}-(\lambda +2 \mu ) U_{0xx} - \varepsilon \Psi_2(U_0, V_1)
+ \frac{\delta^2}{8}\bigg[\frac{\rho^2 c^4}{\mu}U_{0tttt} - \frac{\rho c^2\left(\lambda^2 + 7\lambda\mu + 8\mu^2\right)}{\mu(\lambda + 2\mu)} U_{0xxtt} \\
+ (3\lambda + 4\mu)U_{0xxxx} + 2(3\lambda + 2\mu)V_{1xxx} - \frac{2 \rho  c^2 \left(\lambda ^2+4 \lambda  \mu +2 \mu ^2\right)}{\mu(\lambda + 2\mu)} V_{1xtt} \bigg]\\
+ O(\varepsilon^2, \varepsilon\delta^2, \delta^4)
= \frac{2\mu(3\lambda + 2\mu)}{\lambda + \mu} T,
\end{split}
\end{equation}
where the nonlinear terms are given by 
\begin{eqnarray*}
\Psi_1 &=& (4l + 2m + 3\lambda + 3\mu) V_1^2 + (4l - 2m + n + \lambda) U_{0x} V_1 + \frac{1}{2} (2l + \lambda) U_{0x}^2, \\
\Psi_2 &=& \left((4l - 2m + n + \lambda) V_1^2 + 2(2l + \lambda) U_{0x} V_1 + \frac12(2l + 4m + 3\lambda + 6\mu) U_{0x}^2 \right)_x.
\end{eqnarray*}
We note that when $\varepsilon = 0$, the equations \eqref{bc_rr_subst} and \eqref{bc_rx_subst} reduce to the equations obtained for the linear problem in~\cite{bostrm2000}.  This coupled system constitutes a rather complicated model, and therefore we aim to obtain a single Boussinesq-type equation, i.e. a simpler two-directional model.

There are two natural ways of deriving a Boussinesq-type model. Firstly, elimination of the function $V_1$ from the equations \eqref{bc_rr_subst} and \eqref{bc_rx_subst} can be made using the asymptotic expression which follows from the equation \eqref{bc_rr_subst}:
\begin{equation} \label{v1_asympt}
V_1(x, t) = \frac{ \mu(3\lambda + 2\mu) P - \lambda(\lambda + \mu) U_{0x}}{2(\lambda + \mu)^2} + \varepsilon f(x,t) + \delta^2 g(x,t) + O(\varepsilon^2, \varepsilon\delta^2, \delta^4),
\end{equation}
where the unknown functions $f$ and $g$ can be found by equating to zero coefficients of $\varepsilon$ and $\delta^2$ in \eqref{bc_rr_subst}. Then substitution of $V_1$ into \eqref{bc_rx_subst} results in the following equation for $U_0$:
\begin{eqnarray} \label{eq_u0_asympt}
\begin{split}
\rho c^2 U_{0tt} - \frac{\mu(3\lambda + 2\mu)}{\lambda + \mu}\left(U_{0xx} + \frac{\lambda}{\lambda + \mu} P_x + 2T\right)
- \varepsilon \left(\gamma_1 U_{0x}^2 + \gamma_2 U_x P + \gamma_3 P^2 \right)_x + \delta^2 \bigg[\frac{\rho ^2 c^4 U_{0tttt}}{8\mu} \\
+ \frac{\mu (3\lambda + 2\mu)^2 U_{0xxxx}}{8(\lambda + \mu)^2} - \frac{\rho c^2 \left(7\lambda^2 + 10\lambda\mu + 4\mu^2\right) U_{0xxtt}}{8(\lambda + \mu)^2} + F \bigg] + O(\varepsilon^2, \varepsilon\delta^2, \delta^4) = 0.
\end{split}
\end{eqnarray}
Here, the nonlinearity coefficients $\gamma_i$ and the function $F$ are given by
\begin{eqnarray*} 
\gamma_1 &=& \frac{3n(\lambda + \mu)\lambda^2 + 2\mu \left[9\lambda^3 + 24\mu\lambda^2 + 21\mu^2\lambda + m(3\lambda + 2\mu)^2 + 2\mu^2 (l + 3\mu)\right]}{4 (\lambda + \mu)^3},\\
\gamma_2 &=& \frac{\left[3\lambda^3 + 5\lambda^2\mu + 2\lambda\mu^2 + 4l\mu^2 + 2\lambda m(3\lambda + 2\mu) - 2\lambda n (\lambda + \mu)\right] \mu(3\lambda + 2\mu)}{2(\lambda + \mu)^4},\\
\gamma_3 &=& \frac{\left[n (\lambda +\mu )-2 \left(\lambda ^2+\lambda  \mu -2 l \mu \right)-2 m (2 \lambda +\mu )\right] \mu^2 (3\lambda + 2\mu)^2}{4(\lambda + \mu)^5},\\
F &=& \frac{3\lambda + 2\mu}{8\mu(\lambda + \mu)^3}\left[ \mu (4\lambda^2 + 5\lambda\mu + 2\mu^2) P_{xxx} - \rho c^2(\lambda^2 + \lambda\mu + \mu^2) P_{xtt}\right].
\end{eqnarray*}

Secondly, another approach is to eliminate the function $V_1$ from \eqref{bc_rr_subst} and \eqref{bc_rx_subst} in the same way as it was done in~\cite{bostrm2000} for the linear problem. In the linear case this second approach avoids using the asymptotic expression \eqref{v1_asympt} and results in the equation of the same type as \eqref{eq_u0_asympt}, but with different dispersive coefficients. Indeed, equations \eqref{bc_rr_subst} and \eqref{bc_rx_subst} can be written in the form
\begin{eqnarray}
L_{1} V_1 + \varepsilon N_1(U_0, V_1, \dots) &=& M_1(U_0, P, \dots) + O(\varepsilon^2, \varepsilon\delta^2, \delta^4),\\
L_{2} V_1 + \varepsilon N_2(U_0, V_1, \dots) &=& M_2(U_0, T, \dots) + O(\varepsilon^2, \varepsilon\delta^2, \delta^4),
\end{eqnarray}
where $L_{1}$ and $L_{2}$ are linear differential operators with constant coefficients acting on $V_1$, while $N_1(U_0, V_1, \dots)$, $M_1 (U_0, P, \dots)$ and $N_2(U_0, V_1, \dots)$, $M_2 (U_0, T, \dots)$ are nonlinear functions of their arguments in the equations \eqref{bc_rr_subst} and \eqref{bc_rx_subst}, respectively. Now, applying $L_{2}$ to the first equation, $L_{1}$ to the second equation and subtracting one from another yields:
\begin{equation}
\varepsilon [L_{2}N_1(U_0, V_1, \dots) - L_{1}N_2(U_0, V_1, \dots)] = L_{2}M_1(U_0, P, \dots) - L_{1}M_2(U_0, T, \dots) + O(\varepsilon^2, \varepsilon\delta^2, \delta^4).
\end{equation}
Here $V_1$ is eliminated from the linear part of the equation exactly, rather than asymptotically. To eliminate it from the nonlinear part we again use the expansion \eqref{v1_asympt} and obtain the following  equation
\begin{equation} \label{eq_u0_bostr}
\begin{split}
 \rho c^2 U_{0tt} - \frac{\mu(3\lambda + 2\mu)}{\lambda + \mu} \left(U_{0xx} + \frac{\lambda}{\lambda + \mu} P_x + 2T\right)
- \varepsilon \left(\gamma_1 U_{0x}^2 + \gamma_2 U_x P + \gamma_3 P^2 \right)_x \\
 + \delta^2 \bigg[\frac{\rho ^2 c^4 (\lambda^2 + 5\lambda\mu + 5\mu^2) U_{0tttt}}{8\mu(\lambda+2\mu)(\lambda+\mu)} - \frac{\rho c^2 \left(6\lambda^2 + 21\lambda \mu + 14\mu^2\right) U_{0xxtt}}{8(\lambda + 2\mu)(\lambda + \mu)} \\
 + \frac{\mu(3\lambda + 2\mu) U_{0xxxx}}{4(\lambda + \mu)} + G\bigg] + O(\varepsilon^2, \varepsilon\delta^2, \delta^4) = 0,
\end{split} 
\end{equation}
where
\begin{equation*} \label{G}
G  = \frac{\mu(3\lambda + 2\mu)}{8(\lambda + \mu)^2} \left[(3\lambda + 2\mu) P_{xxx} - \frac{\rho c^2(\lambda^2 + 4\lambda\mu + 2\mu^2)}{\mu(\lambda + 2\mu)} P_{xtt} - \frac{2\rho c^2 (2\lambda + 3\mu)}{\lambda + 2\mu} T_{tt} - 2\lambda T_{xx}\right]. 
\end{equation*}
We note that in the linear approximation, when $\varepsilon = 0$, equation \eqref{eq_u0_bostr} reduces to the equation derived for the linear problem in~\cite{bostrm2000}.

From \eqref{eq_u0_asympt} and \eqref{eq_u0_bostr}, letting $\varepsilon = 0$, $\delta = 0$ and $P = T = 0$, we recover the linear longitudinal wave speed for a thin rod 
\begin{equation}
\label{lin_velocity}
c = \ \sqrt{\frac{\mu(3\lambda + 2\mu)}{\rho(\lambda+\mu)}} = \sqrt{\frac{E}{\rho}}.
\end{equation}
We now rewrite both derived Boussinesq-type equations \eqref{eq_u0_asympt} and \eqref{eq_u0_bostr} in the unified form where all coefficients are given in terms of the Young modulus $ E $ and the Poisson ratio $ \nu $:
\begin{equation}\label{eq_u0_fin}
\begin{split}
U_{0tt} - U_{0xx} &- 2\left(\nu P_{x} + T\right) - \frac{\varepsilon}{2 E} \left(\beta_1U_{0x}^2 + 2 \beta_2 U_{0x} P + \beta_3 P^2 \right)_x \\
&+ \delta^2 \left(\alpha_1^{(i)} U_{0tttt} + \alpha_2^{(i)} U_{0xxtt} + \alpha_3^{(i)} U_{0xxxx} + F^{(i)}\right) + O(\varepsilon^2, \varepsilon\delta^2, \delta^4) = 0, \quad i = 1,2,
\end{split}
\end{equation}
where
\begin{eqnarray*}
\label{alpha1}
&& \alpha_1^{(1)} = \alpha_3^{(1)} = \frac{1 + \nu}{4}, \quad \alpha_2^{(1)} = -\frac{1 + \nu + \nu^2}{2},\\
\label{alpha2}
&& \alpha_1^{(2)} = \frac{5 - 5\nu - 6\nu^2 + 4\nu^3}{8(1-\nu)},\quad \alpha_2^{(2)} = -\frac{7 - 7\nu - 2\nu^2}{8(1-\nu)}, \quad \alpha_3^{(2)} = \frac14,
\end{eqnarray*}
\begin{eqnarray*}
\label{beta_1}
\beta_1 &=& 3E + 2l(1 - 2\nu)^3 + 4m(1 + \nu)^2 (1 - 2\nu) + 6n\nu^2,\\
\label{beta_2}
\beta_2 &=& 2 (1 + \nu) \left[2 l (1 - 2 \nu)^3 + \nu \left(E + 4m \left(1 - \nu - 2\nu^2\right) - 2n (1 - 2\nu)\right) \right],\\
\label{beta_3}
\beta_3 &=& 2(1 + \nu)(1 - 2 \nu) \left[4l \left(1 - 3\nu + 4\nu^3\right) - 2m(1 + \nu)(1 - 4\nu^2) - \nu(2E + 2\nu n + n) + n\right],\\
\label{F1}
F^{(1)} &=& \frac{1}{4} \left[(1 + \nu + 2\nu^2) P_{xxx} - (1 - \nu + 2\nu^2 + 4\nu^3) P_{xtt}\right],\\
\label{F2}
F^{(2)} &=& \frac{1}{4} \bigg[ (1 + \nu) P_{xxx} - \frac{1 + \nu - 2\nu^2 - 2\nu^3}{1 - \nu} P_{xtt} - \frac{3 - 5\nu - 4\nu^2 + 4\nu^3}{2(1 - \nu)} T_{tt} - 2\nu T_{xx}\bigg].
\end{eqnarray*}
Differentiating \eqref{eq_u0_fin} with respect to $x$ we obtain two equations for the longitudinal ``strain" $e = U_{0x}$ in the form
\begin{equation}\label{eq_u0_def_fin}
\begin{split}
e_{tt} - e_{xx} &- 2\left(\nu P_{xx} + T_x\right) - \frac{\varepsilon}{2 E} \left(\beta_1 e^2 + 2 \beta_2 e P + \beta_3 P^2\right)_{xx}\\
& + \delta^2 \left(\alpha_1^{(i)} e_{tttt} + \alpha_2^{(i)} e_{xxtt} + \alpha_3^{(i)} e_{xxxx} + F^{(i)}_x \right) + O(\varepsilon^2, \varepsilon\delta^2, \delta^4) = 0, \quad i = 1,2.
\end{split}
\end{equation}

Three different asymptotic models emerge from the equations \eqref{eq_u0_def_fin} depending on the relative values of the two small parameters present in the problem. 
Firstly, if nonlinearity is very weak compared to dispersion, i.e.  $ \varepsilon\ll\delta^2\ll1 $, we can asymptotically reduce the equations (\ref{eq_u0_def_fin}) to the linear equations
\begin{equation}\label{eq_u0_def_3}
e_{tt} - e_{xx} - 2\left(\nu P_{xx} + T_x\right) + \delta^2 \left(\alpha_1^{(i)} e_{tttt} + \alpha_2^{(i)} e_{xxtt} + \alpha_3^{(i)} e_{xxxx} + F^{(i)}_x \right) + O(\delta^4) = 0, \quad i = 1,2,
\end{equation}
which indicates that in this case the initial evolution will be dominated by dispersion.
Secondly, if nonlinearity is very strong compared to dispersion, i.e.  $ \delta^2\ll\varepsilon\ll1 $, we obtain a non-dispersive equation
\begin{equation}\label{eq_u0_def_2}
e_{tt} - e_{xx} - 2\left(\nu P_{xx} + T_x\right) - \frac{\varepsilon}{2 E} \left(\beta_1 e^2 + 2 \beta_2 e P + \beta_3 P^2\right)_{xx} + O(\varepsilon^2) = 0,
\end{equation}
indicating that in that case the initial evolution will be dominated by nonlinearity.
Finally, if there is a balance between nonlinear and dispersive terms, i.e. $ \varepsilon \sim \delta^2 $, we obtain the ``maximal balance model" (see \cite{Ablowitz} for the terminology):
\begin{equation}\label{eq_u0_def_1}
\begin{split}
e_{tt} - e_{xx} &- 2\left(\nu P_{xx} + T_x\right) - \varepsilon \bigg[\frac{1}{2E} \left(\beta_1 e^2 + 2 \beta_2 e P + \beta_3 P^2\right)_{xx}\\
& + \frac{\delta^2}{\varepsilon} \left(\alpha_1^{(i)} e_{tttt} + \alpha_2^{(i)} e_{xxtt} + \alpha_3^{(i)} e_{xxxx} + F^{(i)}_x \right)\bigg] + O(\varepsilon^2) = 0, \quad i = 1,2.
\end{split}
\end{equation}
The last asymptotic model \eqref{eq_u0_def_1} is a  Boussinesq-type equation (two versions of it), and it is well-known that such equations can support solitary wave solutions or {\it solitons} for brevity (see, for example, \cite{Ablowitz, S_book}). 

We now truncate  the equations \eqref{eq_u0_def_1}  by omitting the terms of order $O(\varepsilon^2)$, and rewrite them in the dimensional form (keeping the same notations for the dimensional variables):
\begin{equation}\label{eq_dim}
\begin{split}
e_{tt} - c^2 e_{xx} &- \frac{2}{\rho}\left(\nu P_{xx} + \frac1R T_x\right) - \left(\frac{\beta_1}{2\rho} e^2 + \frac{\beta_2}{\rho E} e P + \frac{\beta_3}{2\rho E^2} P^2\right)_{xx}\\
& + R^2 \bigg(\frac{\alpha_1^{(i)}}{c^2} e_{tttt} + \alpha_2^{(i)} e_{xxtt} + c^2\alpha_3^{(i)} e_{xxxx} + G^{(i)} \bigg) = 0, \quad i = 1,2,
\end{split}
\end{equation}
where $c^2 = E/\rho$, other constant coefficients are given by
\begin{eqnarray}
\label{alpha1}
&& \alpha_1^{(1)} = \alpha_3^{(1)} = \frac{1 + \nu}{4}, \quad \alpha_2^{(1)} = -\frac{1 + \nu + \nu^2}{2},\\
\label{alpha2}
&& \alpha_1^{(2)} = \frac{5 - 5\nu - 6\nu^2 + 4\nu^3}{8(1-\nu)},\quad \alpha_2^{(2)} = -\frac{7 - 7\nu - 2\nu^2}{8(1-\nu)}, \quad \alpha_3^{(2)} = \frac14,
\end{eqnarray}
\begin{eqnarray}
\label{beta_1}
\beta_1 &=& 3E + 2l(1 - 2\nu)^3 + 4m(1 + \nu)^2 (1 - 2\nu) + 6n\nu^2,\\
\label{beta_2}
\beta_2 &=& 2 (1 + \nu) \left[2 l (1 - 2 \nu)^3 + \nu \left(E + 4m \left(1 - \nu - 2\nu^2\right) - 2n (1 - 2\nu)\right) \right],\\
\label{beta_3}
\beta_3 &=& 2(1 + \nu)(1 - 2 \nu) \left[4l \left(1 - 3\nu + 4\nu^3\right) - 2m(1 + \nu)(1 - 4\nu^2) - \nu(2E + 2\nu n + n) + n\right],
\end{eqnarray}
and the dimensional functions $G^{(i)}$ have the form
\begin{eqnarray}
G^{(1)} &=& \frac{1 + \nu + 2\nu^2}{4\rho} P_{xxxx} - \frac{1 - \nu + 2\nu^2 + 4\nu^3}{4E} P_{xxtt}, \label{G1}\\
G^{(2)} &=& \frac{1 + \nu}{4\rho} P_{xxxx} - \frac{1 + \nu - 2\nu^2 - 2\nu^3}{4E(1 - \nu)} P_{xxtt} - \frac{3 - 5\nu - 4\nu^2 + 4\nu^3}{8ER(1 - \nu)} T_{xtt} - \frac{\nu}{2\rho R} T_{xxx}. \label{G2}
\end{eqnarray}

The equations (\ref{eq_u0_def_1}) have been derived for the case of strong tractions, when the related terms are present at the leading order. If tractions are relatively weak,
$$
P = \varepsilon \hat P, \quad T = \varepsilon \hat T,
$$
then the equations (\ref{eq_u0_def_1}) asymptotically reduce to 
\begin{equation}\label{eq_u0_def_1a}
\nonumber
\begin{split}
e_{tt} - e_{xx}  - \varepsilon \bigg[& 2\left(\nu \hat P_{xx} + \hat T_x\right) + \frac{1}{2E} \left(\beta_1 e^2  \right)_{xx} \\
& + \frac{\delta^2}{\varepsilon} \left(\alpha_1^{(i)} e_{tttt} + \alpha_2^{(i)} e_{xxtt} + \alpha_3^{(i)} e_{xxxx}  \right)\bigg] + O(\varepsilon^2) = 0, \quad i = 1,2,
\end{split}
\end{equation}
and the dimensional form of these equations is given by
\begin{equation}\label{eq_dim_weak_trac}
\begin{split}
e_{tt} - c^2 e_{xx} &- \frac{2}{\rho}\left(\nu P_{xx} + \frac1R T_x\right)  - \left(\frac{\beta_1}{2\rho} e^2 \right)_{xx} \\
& + R^2 \bigg(\frac{\alpha_1^{(i)}}{c^2} e_{tttt} + \alpha_2^{(i)} e_{xxtt} + c^2\alpha_3^{(i)} e_{xxxx}  \bigg) = 0, \quad i = 1,2.
\end{split}
\end{equation}

We note that in case of the free surface boundary conditions, when $P = T = 0$, the equations \eqref{eq_dim} (and  \eqref{eq_dim_weak_trac}) reduce to
\begin{equation}\label{eq_dim_free_surf}
e_{tt} - c^2 e_{xx} = \frac{\beta_1}{2\rho}\left(e^2\right)_{xx} -  R^2 \left(\frac{\alpha_1^{(i)}}{c^2} e_{tttt} + \alpha_2^{(i)} e_{xxtt} + c^2\alpha_3^{(i)} e_{xxxx}\right), \quad i = 1,2.
\end{equation}

It is now instructive to compare both of  equations  in \eqref{eq_dim_free_surf} with the ``doubly dispersive equation (DDE)" derived by A.M. Samsonov and A.V. Porubov \cite{SP, S_book, P_book}:
\begin{equation}\label{eq_dim_SP}
e_{tt} - c^2 e_{xx} =  \frac{\beta_1}{2 \rho} (e^2)_{xx} - \frac{\nu (1-\nu) R^2}{2} e_{xxtt} + \frac{\nu c^2 R^2}{2} e_{xxxx},
\end{equation}
and the equation derived by L.A. Ostrovsky and A.M. Sutin \cite{OS}:
\begin{equation}\label{eq_dim_OS}
e_{tt} - c^2 e_{xx} =  \frac{\beta_1}{2 \rho} (e^2)_{xx} + \frac{\nu^2 R^2}{2} e_{xxtt}.
\end{equation}
Here, all four models have the same nonlinear terms, but they have different dispersive terms. Equations \eqref{eq_dim_SP} and \eqref{eq_dim_OS} can also be written in the form of the equation \eqref{eq_dim_free_surf} with the following dispersive coefficients, respectively:
\begin{equation*}
\begin{split}
&\alpha_1^{(3)} = 0,\quad \alpha_2^{(3)} = \frac{(1-\nu)\nu}{2}, \quad \alpha_3^{(3)} = -\frac \nu 2,\\
&\alpha_1^{(4)} = 0,\quad \alpha_2^{(4)} = -\frac{\nu^2}{2}, \quad \alpha_3^{(4)} = 0.
\end{split}
\end{equation*}

We note that all four Boussinesq-type models shown above are not asymptotically exact equations, i.e. in non-dimensional form they include both $O(1)$ and $O(\varepsilon)$ terms. Hence all these equations can be ``regularised'' to the form where there is just one dispersive term, using the leading order relation $ e_{tt} = c^2 e_{xx} + \dots \;$. Coefficient of that dispersive term is a sum of all dispersive coefficients and it is the same for all four equations:
\begin{equation} \label{alpha_sum}
\alpha_1^{(i)} + \alpha_2^{(i)} + \alpha_3^{(i)} = -\frac{\nu^2}{2}, \quad i = \overline{1,4},
\end{equation}
which means these equations are asymptotically equivalent.

However, prior to the regularisation, the models have different dispersive properties, and, similarly to the linear studies (see \cite{bostrm2000} and references to the classical linear results therein), it would be interesting to compare the performance of these four nonlinear models with the exact (numerical) solution of the nonlinear problem.  We also note that the three dispersive terms present in the equation (\ref{eq_dim_free_surf}), $i = 1$ are similar to the dispersive terms in the Boussinesq-type equation derived from a nonlinear lattice model for a waveguide in \cite{KSZ}. 

We note that regularisation of the type discussed above has been introduced by T.B. Benjamin, J.L. Bona and J.J. Mahony in the context of fluids \cite{BBM}, and has been further developed  in the context of solids by P. Rosenau, and M.B. Rubin, P. Rosenau and O. Gottlieb  \cite{R, RRG}.

\section{Alternative derivation}
In this section we aim to show that equation \eqref{eq_u0_asympt} can also be derived from the full problem formulation \eqref{eq1_0} and \eqref{eq2_0} using the relaxed assumptions on the form of the asymptotic expansions, and therefore justifying the expansions in the form of  the power series in the radial variable used in the previous section. 

Indeed, let us introduce again the dimensionless variables (\ref{scales1}) and look for
dimensionless displacements in the form of asymptotic expansions in the small parameter $\delta$:
\begin{eqnarray}
\label{expand_delta1}
U(x, r, t) &=& U_0(x, t) + U_2(x, r, t) \delta^2 + U_4(x, r, t) \delta^4 + O(\delta^6) ;\\
\label{expand_delta2}
V(x, r, t) &=& V_1(x, r, t) + V_3(x, r, t) \delta^2 + V_5(x, r, t) \delta^4 + O(\delta^6).
\end{eqnarray}
Here, for brevity, we already used that to the leading order $U_0$ is independent of $r$ (plane cross sections), which immediately follows from the leading order (linear) approximation (see \cite{bostrm2000} and references therein). We would like to emphasise  that this is not a simplifying assumption, but a rigorous leading-order result of this asymptotic theory.

Substituting the expansions \eqref{expand_delta1} and \eqref{expand_delta2} into the equations of motion \eqref{eq1_0} and \eqref{eq2_0} we obtain
\begin{eqnarray}\label{eq1_2}
\begin{split}
&\rho c^2 U_{0tt} - (\lambda + 2\mu) U_{0xx} - (\lambda + \mu) \left(V_{1xr} + \frac{V_{1x}}{r}\right) - \mu\left(U_{2rr} + \frac{U_{2r}}{r}\right) + \varepsilon\widetilde{\Phi}_1(U_0, V_1, U_2) \\
&+ \delta^2 \left[\rho c^2 U_{2tt} - (\lambda + 2\mu) U_{2xx} - (\lambda + \mu) \left(V_{3xr} + \frac{V_{3x}}{r}\right) - \mu\left(U_{4rr} + \frac{U_{4r}}{r}\right)\right] + O(\varepsilon^2, \varepsilon\delta^2, \delta^4) = 0,
\end{split} 
\end{eqnarray}
\begin{equation}\label{eq2_2}
\begin{split}
(\lambda + 2\mu) \left(\frac{V_1}{r^2} - \frac{V_{1r}}{r} - V_{1rr}\right) + \varepsilon \widetilde{\Phi}_2(U_0, V_1) &+ \delta^2 \bigg[\rho c^2 V_{1tt} - \mu V_{1xx} - (\lambda + \mu)U_{2xr} \\
& + (\lambda + 2\mu)\left(\frac{V_3}{r^2} - \frac{V_{3r}}{r} - V_{3rr}\right) \bigg] + O(\varepsilon^2, \varepsilon\delta^2, \delta^4) = 0,
\end{split}
\end{equation}
where the functions indicated by $\widetilde{\Phi}_1$ and $\widetilde{\Phi}_2$ include all nonlinear terms (not shown here because they are given by very long expressions). 

The boundary conditions at $r = 1$ have the form
\begin{eqnarray}\label{bc_rr_2}
&&\begin{split}
\lambda U_{0x} + (\lambda + 2\mu)V_{1r} + \lambda V_1 + \varepsilon\widetilde{\Psi}_1(U_0, V_1, U_2, V_3) + \delta^2\big[\lambda  U_{2x} + (\lambda + 2\mu)V_{3r} + \lambda V_3\big]\\
+ O(\varepsilon^2, \varepsilon\delta^2, \delta^4) =  \frac{\mu(3\lambda + 2\mu)}{\lambda + \mu} P(x,t),
\end{split}\\
\label{bc_rx_2}
&&\mu \left(U_{2r} + V_{1x}\right) + \varepsilon\widetilde{\Psi}_2(U_0, V_1, U_2, V_3) + \delta^2 \mu \left(U_{4r}+V_{3x}\right) + O(\varepsilon^2, \varepsilon\delta^2, \delta^4) = \frac{\mu(3\lambda + 2\mu)}{\lambda + \mu} T(x,t).
\end{eqnarray}
Collecting the coefficients in front of the equal powers of $\delta$ in the equations \eqref{eq1_2} and \eqref{eq2_2} we obtain a set of nonlinear ordinary differential equations for the functions of variable $r$, where all nonlinear terms are multiplied by $\varepsilon$. We solve these equations with the boundary conditions following from  \eqref{bc_rr_2} and \eqref{bc_rx_2} using asymptotic expansions of functions in $\varepsilon$. For example, we write the function $V_1$ as
\begin{equation}\label{v1_eps_expantion}
V_1(x,r,t) = f(x,r,t) + \varepsilon g(x,r,t) + \O{\varepsilon^2},
\end{equation}
where $f$ and $g$ are some unknown functions. Substituting \eqref{v1_eps_expantion} into \eqref{eq2_2} we obtain an ODE for the function $f$ at the leading order of $\varepsilon$:
\begin{equation}\label{v1_f_ode}
f_{rr} + \frac{f_r}{r} - \frac{f}{r^2} = 0.
\end{equation}
The boundary condition following from \eqref{bc_rr_2} takes the form
\begin{equation}\label{v1_f_bc1}
(\lambda + 2\mu)f_r + \lambda f = -\lambda U_{0x} + \frac{\mu(3\lambda + 2\mu)}{\lambda + \mu} P(x,t) \quad\text{at } r = 1.
\end{equation}
The problem \eqref{v1_f_ode}, \eqref{v1_f_bc1} is complemented by the symmetry condition requiring that the radial displacement at the centre of the rod is equal to zero:
\begin{equation}\label{v1_f_bc2}
f = 0 \quad\text{at } r = 0.
\end{equation}
The general solution of the equation \eqref{v1_f_ode} has the form
\begin{equation}\label{v1_f_gen_sln}
f(x,r,t) = C_1(x,t)r + \frac{C_2(x,t)}{r} .
\end{equation}
From \eqref{v1_f_bc2} it follows that $C_2 \equiv 0$, and $C_1$ is obtained from \eqref{v1_f_bc1}, yielding
\begin{equation}\label{v1_f_sln}
f(x,r,t) = \frac{r}{2(\lambda + \mu)} \left(\frac{ \mu(3\lambda + 2\mu)}{\lambda + \mu} P - \lambda U_{0x}\right).
\end{equation}
Using \eqref{v1_f_sln},  we then obtain an equation for the function $g$ at the next order of $\varepsilon$ as
\begin{equation}\label{v1_g_ode}
g_{rr} + \frac{g_r}{r} - \frac{g}{r^2} = 0,
\end{equation}
with the boundary conditions 
\begin{equation}\label{v1_g_bc1}
(\lambda + 2\mu)g_r + \lambda g = a_1 U_{0x}^2 + a_2 U_{0x} P + a_3 P^2 \quad\text{at } r = 1,
\end{equation}
\begin{equation}\label{v1_g_bc2}
g = 0 \quad\text{at } r = 0.
\end{equation}
The solution of \eqref{v1_g_ode}, \eqref{v1_g_bc1}, \eqref{v1_g_bc2} is given by
\begin{equation}\label{v1_g_sln}
g(x,t) = \frac{r (a_1 U_{0x}^2 + a_2 U_{0x} P + a_3 P^2)}{2(\lambda + \mu)}.
\end{equation}
We eliminate functions $U_2$, $V_3$ and $U_4$ in a similar way, imposing the symmetry condition 
$U_r = 0$ at $r = 0$. 

Finally we obtain
\begin{eqnarray}
&&V_1(x, r, t) = \frac{r}{2(\lambda + \mu)} \left(\frac{ \mu(3\lambda + 2\mu)}{\lambda + \mu} P - \lambda U_{0x}  + \varepsilon (a_1 U_{0x}^2 + a_2 U_{0x} P + a_3 P^2) \right) + O(\varepsilon^2),\\
&&\begin{split}
U_2(x, r, t) = &\frac{r^2}{4\mu} \left(\rho c^2 U_{0tt} - 2\mu U_{0xx} - \frac{\mu(3\lambda + 2\mu)}{\lambda + \mu} P_x\right)\\
& + \varepsilon r^2 \big[U_{0x} \left(a_4 U_{0tt} + a_5 U_{0xx} + a_6 P_x\right) + P(a_7 U_{0tt} + a_8 U_{0xx} + a_9 P_x)\big] + O(\varepsilon^2),
\end{split}\\
&&V_3(x, r, t) = r \left(b_1(r) P_{tt} + b_2(r) U_{0xtt} + b_3(r) P_{xx} + b_4(r) U_{0xxx}\right) + O(\varepsilon),\\
&&U_4(x, r, t) = r^4 a_{10} U_{0tttt} + r^2 \left(b_5(r) U_{0xxxx} + b_6(r) U_{0xxtt} + b_7(r) P_{xtt} + b_8(r) P_{xxx}\right) + O(\varepsilon).
\end{eqnarray}
Here, $b_i(r) = b_{i}^{(2)}r^2 + b_{i}^{(0)}$, the coefficients $a_i$ and $b_{i}^{(j)}$ depend on the elastic moduli $\lambda, \mu, l, m, n$ and the density $\rho$.

The resulting equation which follows from the equation \eqref{bc_rx_2} after the substitution of $V_1$, $U_2$, $V_3$ and $U_4$ coincides with the previously derived equation~\eqref{eq_u0_asympt}. 

\section{Derivation of a Boussinesq-type equation for a pre-stretched rod}
In this section we consider the propagation of longitudinal waves in the uniformly pre-stretched rod (in the axial direction). The longitudinal displacement in the pre-stretched state  is given by
\begin{equation}\label{pre_stretch_u}
U^*(x) = \kappa x,
\end{equation}
where $\kappa$ is the constant longitudinal pre-strain.
We non-dimensionalise the pre-stretch using the same scaling factor as $U$ in \eqref{scales1} which yields 
\begin{equation} \label{scales_pre_stretch}
\tilde U^* = \frac{U^*}{\varepsilon L} = \tilde\kappa \tilde x, \quad \mbox{where} \quad  \tilde \kappa = \frac{\kappa}{\varepsilon}.
\end{equation}
Moreover, we assume that in the initial pre-stretched state there are zero tractions on the rod's lateral surface. Solving the equations of motion \eqref{eq1_0} and \eqref{eq2_0} with the free surface (i.e. $P = T = 0$)  boundary conditions \eqref{bc_rr} and \eqref{bc_rx} written in dimensionless form using \eqref{scales1} and \eqref{scales_pre_stretch}, we obtain the radial displacement $\tilde V^*$ in the pre-stretched rod: 
\begin{equation}\label{pre_stretch_v}
\tilde V^*(\tilde r) = -\,\frac{\lambda}{2(\lambda+\mu)}\tilde\kappa \tilde r \left(1 + \varepsilon \frac{\tilde\kappa \left(2\mu^2 (\lambda + 2l) + \lambda^2 (3\lambda + 6m - 2n) + \lambda\mu(5\lambda + 4m - 2n)\right)}{4\lambda(\lambda + \mu)^2} + O(\varepsilon^2)\right).
\end{equation}
We introduce new dimensionless power series expansions of displacements (tildes omitted):
\begin{eqnarray}
\label{pre_stretch_u_series_scaled}
U(x,r,t) &=& \varepsilon L \left(U^*(x) + U_0 + \delta^2 r^2 U_2 + \delta^4 r^4 U_4 + O(\delta^6)\right),\\
\label{pre_stretch_v_series_scaled}
V(x,r,t) &=& \varepsilon L \delta \left(V^*(r) + r V_1 + \delta^2 r^3 V_3 + \delta^4 r^5 V_5 + O(\delta^6)\right).
\end{eqnarray}
Following the steps of the derivation of the equations \eqref{eq_u0_def_fin} using the power series \eqref{pre_stretch_u_series_scaled}, \eqref{pre_stretch_v_series_scaled} instead of \eqref{u_series_scaled}, \eqref{v_series_scaled} we obtain the equations
\begin{equation}\label{pre_stretch_eq_u0_def_fin}
\begin{split}
e_{tt} - \left(1 + \varepsilon\kappa\frac{\beta_1}{E}\right)e_{xx} - 2 \left[\left(\nu + \varepsilon\kappa \frac{\beta_2}{2E} \right) P_{xx} + T_x\right] - \varepsilon \left(\frac{\beta_1}{2 E} e^2 + \frac{\beta_2}{E} e P + \frac{\beta_3}{2E} P^2\right)_{xx}\\
+ \delta^2 \left(\alpha_1^{(i)} e_{tttt} + \alpha_2^{(i)} e_{xxtt} + \alpha_3^{(i)} e_{xxxx} + F^{(i)}_x \right) + O(\varepsilon^2, \varepsilon\delta^2, \delta^4) = 0, \quad i = 1,2,
\end{split}
\end{equation}
where we used notations introduced in the previous sections.
We note that here $e = U_{0x}$ is the deviation from the pre-stretched state, while in the equations \eqref{eq_u0_def_fin} it represents the deviation from the undeformed state.

Assuming that nonlinear and dispersive terms are of the same order ($\varepsilon \sim \delta^2$) and truncating \eqref{pre_stretch_eq_u0_def_fin} we obtain the equation, which in dimensional variables takes the form
\begin{equation}\label{pre_stretch_eq_dim}
\begin{split}
e_{tt} - \left(c^2 + \kappa\frac{\beta_1}{\rho}\right) e_{xx} - \frac{2}{\rho}\left[\left(\nu + \kappa \frac{\beta_2}{2E} \right) P_{xx} + \frac1R T_x\right] - \left(\frac{\beta_1}{2\rho} e^2 + \frac{\beta_2}{\rho E} e P + \frac{\beta_3}{2\rho E^2} P^2\right)_{xx}\\
+ R^2 \left(\frac{\alpha_1^{(i)}}{c^2} e_{tttt} + \alpha_2^{(i)} e_{xxtt} + c^2\alpha_3^{(i)} e_{xxxx} + G^{(i)} \right) = 0, \quad i = 1,2,
\end{split}
\end{equation}
where the coefficients $\alpha_j^{(i)}$, $\beta_j$ and the functions $G^{(i)}$ are given by the formulae \eqref{alpha1} -- \eqref{beta_3} and (\ref{G1}) -- (\ref{G2}), respectively.

In the case of the weak tractions discussed in Section 3 this equation asymptotically reduces to
\begin{eqnarray}\label{pre_stretch_weak_trac_eq_dim}
\begin{split}
e_{tt} - \left(c^2 + \kappa\frac{\beta_1}{\rho}\right) e_{xx} &- \frac{2}{\rho}\left[\left(\nu + \kappa \frac{\beta_2}{2E} \right) P_{xx} + \frac1R T_x\right] - \left(\frac{\beta_1}{2\rho} e^2 \right)_{xx}\\
&+ R^2 \left(\frac{\alpha_1^{(i)}}{c^2} e_{tttt} + \alpha_2^{(i)} e_{xxtt} + c^2\alpha_3^{(i)} e_{xxxx} \right) = 0, \quad i = 1,2.
\end{split}
\end{eqnarray}

We note that the acoustoelastic effect (modification of the linear wave speed in a pre-stressed media) has been studied in \cite{HughesKelly, ADO, P} (see also references therein). To the best of our knowledge, both models derrived in our paper and described by the equations (\ref{pre_stretch_eq_dim}) (as well as their reduced versions (\ref{eq_dim}) and (\ref{eq_dim_free_surf})) have not been obtained before.

\section{Dispersive properties and solitary wave solutions}

In Fig. \ref{fig:disp} we compare the linear dispersion curves of the four basic (i.e. with the free surface and no pre-stretch) Boussinesq-type equations listed in the previous section, as well as plotting the three lowest branches of the exact (Pochhammer - Chree) dispersion relation of the linear problem for a circular rod (see, for example, \cite{DF, bostrm2000, Love}). 

The dispersive relations have the following form:
\begin{eqnarray}
&&\frac{2p}{R}\left(q^2 + k^2\right) J_1(pR) J_1(qR) - \left(q^2 - k^2\right)^2 J_0(pR) J_1(qR) -4k^2 p q J_1(pR) J_0(qR) = 0,\\
&& \alpha_1^{(i)} \overline{\omega}^4 - \left(1 - \alpha_2^{(i)} \overline{k}^2\right) \overline{\omega}^2 + \overline{k}^2 \left(1 + \alpha_3^{(i)} \overline{k}^2 \right) = 0, \quad i = 1,2,\\
&& \left(1 - \frac{(1-\nu)\nu}{2} \overline{k}^2\right) \overline{\omega}^2 - \overline{k}^2 \left(1 - \frac{\nu  \overline{k}^2}{2}\right) = 0,\\
&& \left(1 + \frac{\nu^2}{2} \overline{k}^2\right) \overline{\omega}^2 - \overline{k}^2 = 0,
\end{eqnarray}
for the Pochhammer - Chree solution and equations (\ref{eq_dim_free_surf}) $i=1,2$, \eqref{eq_dim_SP}  and \eqref{eq_dim_OS} respectively. Here $\overline{k} = k R$, $\overline{\omega} = \omega R / c$, $k$ and $\omega$ are the wavenumber and wave frequency respectively, $J_i$ are the Bessel functions of the first kind, and parameters $p$ and $q$ are given by
\begin{equation}
p^2 = \frac{\rho \omega^2}{\lambda + 2\mu} - k^2, \quad q^2 = \frac{\rho \omega^2}{\mu} - k^2.
\end{equation}
All models reasonably well describe the lowest branch of the dispersion curves for the long waves. Eq. (\ref{eq_dim_SP})  suffers from a short-wave instability, while other three models do not have this defect. Eq. (\ref{eq_dim_free_surf}) for $i=1,2$, capture the presence of the second branch. We also note that, at least in this example, eq. (\ref{eq_dim_free_surf}) for $i=1$ has better dispersive properties than eq. (\ref{eq_dim_free_surf}) for $i=2$ (as a long-wave model). However, eq. (\ref{eq_dim_free_surf}) for $i=2$ better describes the lowest branch in the short wave region. One can expect that both derived Boussinesq-type models in (\ref{eq_dim_free_surf}) can be useful, depending on the type of the dominant dispersive radiation in the problem under study. One could also try to artificially ``optimise" the dispersive properties as discussed, for example, in \cite{PAT, ADKM}. However, in this paper we are interested in the ``natural" derivation of Boussinesq-type models.

\begin{figure}[h]
	\centering
	\includegraphics[width=0.75\linewidth]{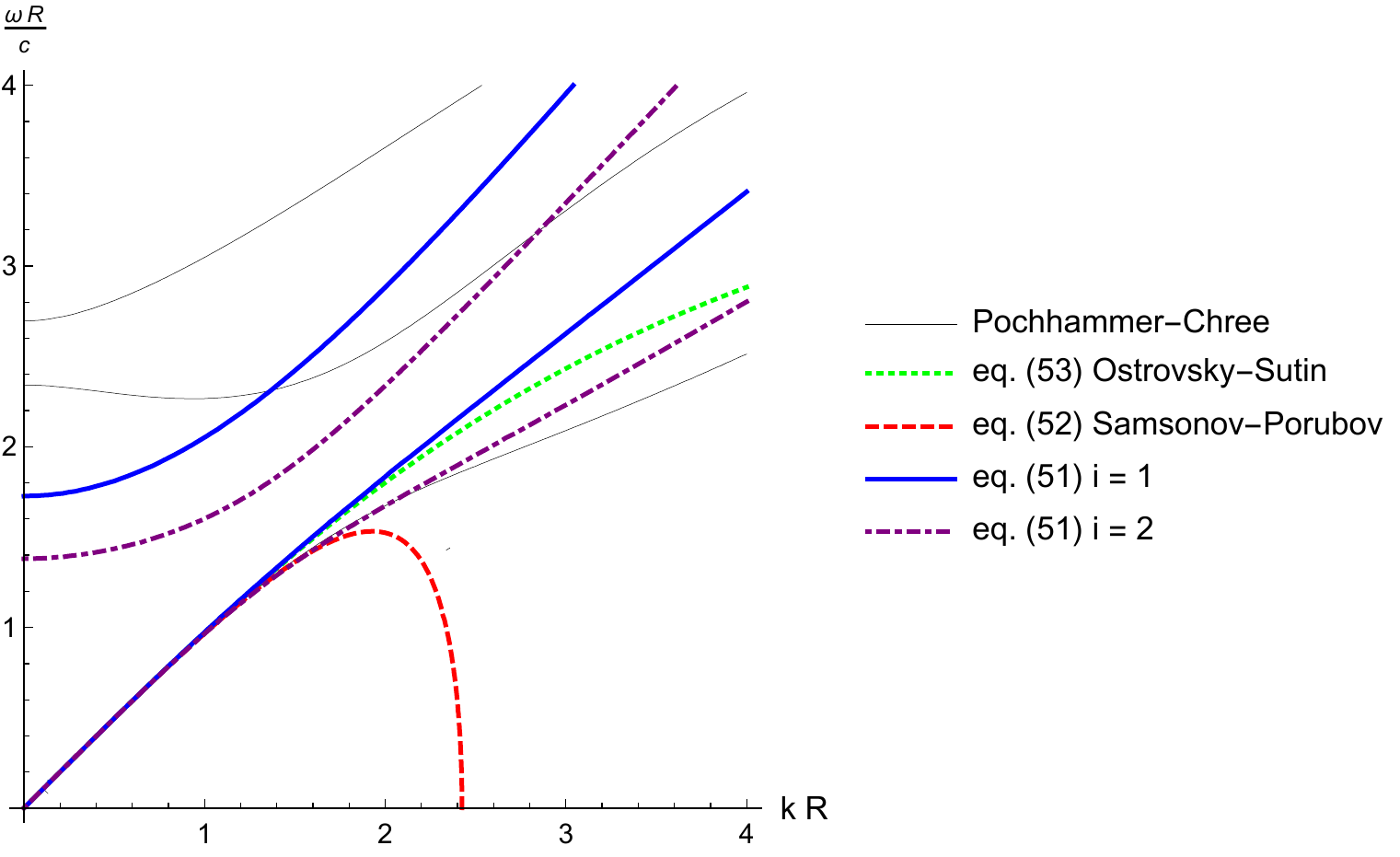}
	\caption{Linear dispersion curves for a rod made of polystyrene (PS) for $R = 10^{-2} m$. PS elastic moduli are given in the Table~\ref{tab:ps}. 
	}
	\label{fig:disp}
\end{figure}

All four equations, i.e. eq. (\ref{eq_dim_free_surf}) for $i=1,2$, eq. \eqref{eq_dim_SP}  and eq. \eqref{eq_dim_OS} have one-parameter families of solitary wave solutions (see Appendix A):
\begin{equation}\label{soliton1}
e_i(x,t) = A\ {\rm sech}^2\ \left[B_{i} \left(x\pm t \sqrt{c^2+\frac{A \beta_1}{3 \rho}}\right) \right], \quad i = \overline {1,4},
\end{equation}
here parametrised by the amplitude $A$. For a given amplitude $A$, the respective solitary wave solutions have the same velocity but different width parameters $B_i$:
\begin{eqnarray}
\label{Bi}
B_i &=& \sqrt{\frac{3A\beta_1 E}{-4\left[(A\beta_1 + 3E)^2\alpha_1^{(i)} + 3E(A\beta_1 + 3E)\alpha_2^{(i)} + 9E^2\alpha_3^{(i)}\right] R^2}} \, , \quad i=1,2,\\
\label{B3}
B_3 &=& \sqrt{\frac{A\beta_1}{\left[6\nu E + 2 A \beta_1 (\nu - 1)\right] \nu R^2}}\, , \\
\label{B4}
B_4 &=& \sqrt{\frac{A\beta_1}{(6E + 2A\beta_1)\nu^2 R^2}},
\end{eqnarray}
for the equations (\ref{eq_dim_free_surf}), 
\eqref{eq_dim_SP} and (\ref{eq_dim_OS})
respectively.

In the left part of the Figure \ref{fig:soliton} we plot the four solitons given by the formulae (\ref{soliton1}) - (\ref{B4}) for one and the same value of the amplitude parameter $A = -0.05$ and the same elastic moduli shown in Table \ref{tab:ps} (typical for a polystyrene \cite{HughesKelly}). We can see that the four solitons have a different width, with the regularised soliton (\ref{B4}) being the widest. However, this figure is plotted for the value of $A$ which exceeds the yield point for the polystyrene, and therefore in practice this difference would not be important for that particular material (but could be important for some other materials). Indeed, in experiments with polystyrene discussed in the next section the value of $A$ is very small, $A \sim 10^{-3} - 10^{-4}$. Therefore, to leading order in $A$, all four formulae will give the width parameter approximately equal to 
\begin{equation}\label{}
B = \sqrt{\frac{A\beta_1}{6\nu^2 E R^2}},
\end{equation}
and the respective solitary wave solution is plotted in the right part of the same  Figure \ref{fig:soliton} for $A = -0.001$.
\begin{figure}[h]
	\centering
	\includegraphics[width=0.54\linewidth]{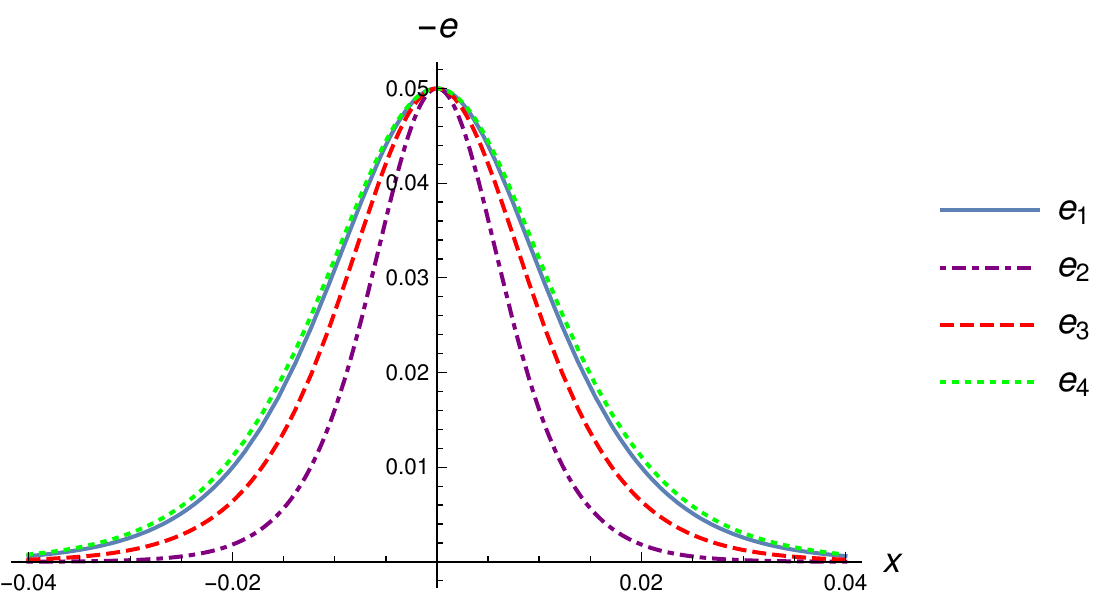}
	\includegraphics[width=0.42\linewidth]{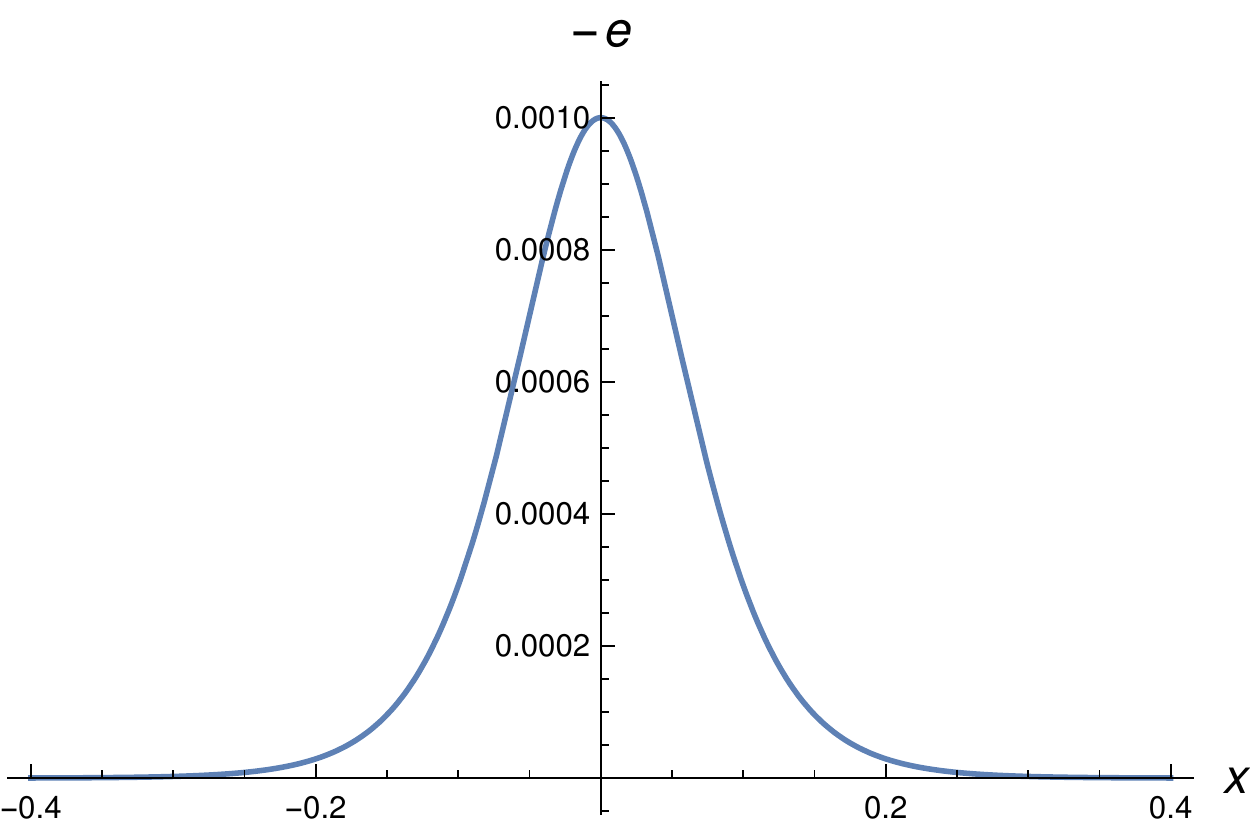}
	\caption{Solitons in a rod made of polystyrene (PS) for $R = 10^{-2} m$, $A = -0.05$ (left) and $A= - 0.001$ (right). PS elastic moduli are given in the Table \ref{tab:ps}. 
	}
	\label{fig:soliton}
\end{figure}

\begin{table}[h]
	\begin{center}
		\begin{tabular}{|c|c|c|c|c|c|}
			\hline
			Young modulus & Poisson & \multicolumn{3}{|c|} {Murnaghan moduli, N/m\textsuperscript{2} } & Density \\
			\cline{3-5}
			$E$, N/m\textsuperscript{2} & ratio, $\nu$ & $l$ & $m$ & $n$ & $\rho$, kg/m\textsuperscript{3}  \\
			\hline
			$3.7\cdot10^9$ & $0.34$ & $-18.9\cdot10^{9}$ & $-13.3\cdot10^{9}$ & $-10\cdot10^{9}$ & 1060 \\
			\hline
		\end{tabular}
	\end{center}
	\caption{Elastic moduli of the polystyrene (PS) \cite{HughesKelly}.}
	\label{tab:ps}
\end{table}

Let us now reparametrise the solitary wave solution \eqref{soliton} by the speed of the soliton $v$ instead of its amplitude $A$:
\begin{equation}
e_i(x, t) = \frac{3 \rho \left(v^2 - c^2\right)}{\beta_1} {\rm sech}^{2}\left[\tilde{B_i}(x\pm v t)\right], \qquad v = \sqrt{c^2+\frac{A\beta_1}{3 \rho}} \, ,
\end{equation}
where
\begin{eqnarray}
\label{Bimod}
\tilde{B}_i &=& \sqrt{\frac{c^2(v^2- c^2)}{-4\left(\alpha_1^{(i)} v^4 + \alpha_2^{(i)} c^2 v^2 + \alpha_3^{(i)}c^4\right)R^2}} \, , \quad i =1,2,\\
\label{B3mod}
\tilde{B}_3 &=& \sqrt{\frac{v^2- c^2}{2\nu R^2 [c^2 - (1 - \nu) v^2 ]}} \, ,\\
\label{B4mod}
\tilde{B}_4 &=& \sqrt{\frac{v^2- c^2}{2\nu^2 v^2 R^2}} \, .
\end{eqnarray}
The solitary wave solution exists only if the width parameter $\tilde{B}$ is real and therefore $\tilde{B}^2 > 0$, which yields, assuming that $\nu < 1$, the following restrictions on the speed of solitary waves:
\begin{itemize}
	\item $\tilde{B}_i^2 > 0 \implies$ $\displaystyle v^2 < \frac{-\alpha_2^{(i)} - \sqrt{\alpha_2^{(i)2} - 4\alpha_1^{(i)}\alpha_3^{(i)}}}{2\alpha_1^{(i)}} c^2$ or $\displaystyle c^2 < v^2 < \frac{-\alpha_2^{(i)} + \sqrt{\alpha_2^{(i)2} - 4\alpha_1^{(i)}\alpha_3^{(i)}}}{2\alpha_1^{(i)}} c^2$, $i=1,2$,
	\item $\tilde{B}_3^2 > 0 \implies$ $\displaystyle c^2 < v^2 < \frac{c^2}{1-\nu}$, 
	\item $\tilde{B}_4^2 > 0 \implies$ $c^2 < v^2$.
\end{itemize}
We also note that $\displaystyle 0 < \frac{-\alpha_2^{(i)} - \sqrt{\alpha_2^{(i)2} - 4\alpha_1^{(i)}\alpha_3^{(i)}}}{2\alpha_1^{(i)}} \leqslant 1$ and $\displaystyle \frac{-\alpha_2^{(i)} + \sqrt{\alpha_2^{(i)2} - 4\alpha_1^{(i)}\alpha_3^{(i)}}}{2\alpha_1^{(i)}} \geqslant 1$ $\forall \nu\in[0, 0.5]$ for $i=1,2$.

Thus, while the first three model equations give a finite range for the speeds of compression solitary waves, the regularised model does not impose an upper bound (see also the related discussions in  \cite{S_book}). Also, the first two models allow for the existence of solitons of opposite polarity, while the other two models do not allow that.   It would be interesting to compare the predictions for the permissible range of soliton speeds and polarities with direct numerical simulations of the full problem formulation. This could guide some future laboratory experiments.

\section{Experimental observations of generation of a soliton}

Experiments on bulk strain soliton generation and monitoring are being performed by the experimental group in the Ioffe Institute in St.Petersburg, Russia for more than 30 years, with the first report on the successful generation of a strain soliton in a polystyrene rod being dated by 1988 \cite{TP1988}. Since then bulk strain solitons were generated and detected, using optical methods, in various waveguides made of three glassy polymers: polystyrene, polymethyl methacrylate and polycarbonate; and in layered waveguides made of combinations of these materials, see, for example, \cite{Dreiden-TP-2008,TPL-Polycarb,JAP2008,JAP2010,JAP2012,APL2014,APL2018}.

Experimental evaluation of various mechanisms of soliton formation has led to the generation triggered by a shock wave formed in water in the vicinity of a waveguide input cross section. The shock wave is formed by evaporation of a metallic foil, using a pulsed laser. Schematic of the experimental setup for soliton generation and recording is shown in Fig. \ref{Excitation_setup}.
\begin{figure}[htbp]
	\centering
	\fbox{\includegraphics[width=8cm]{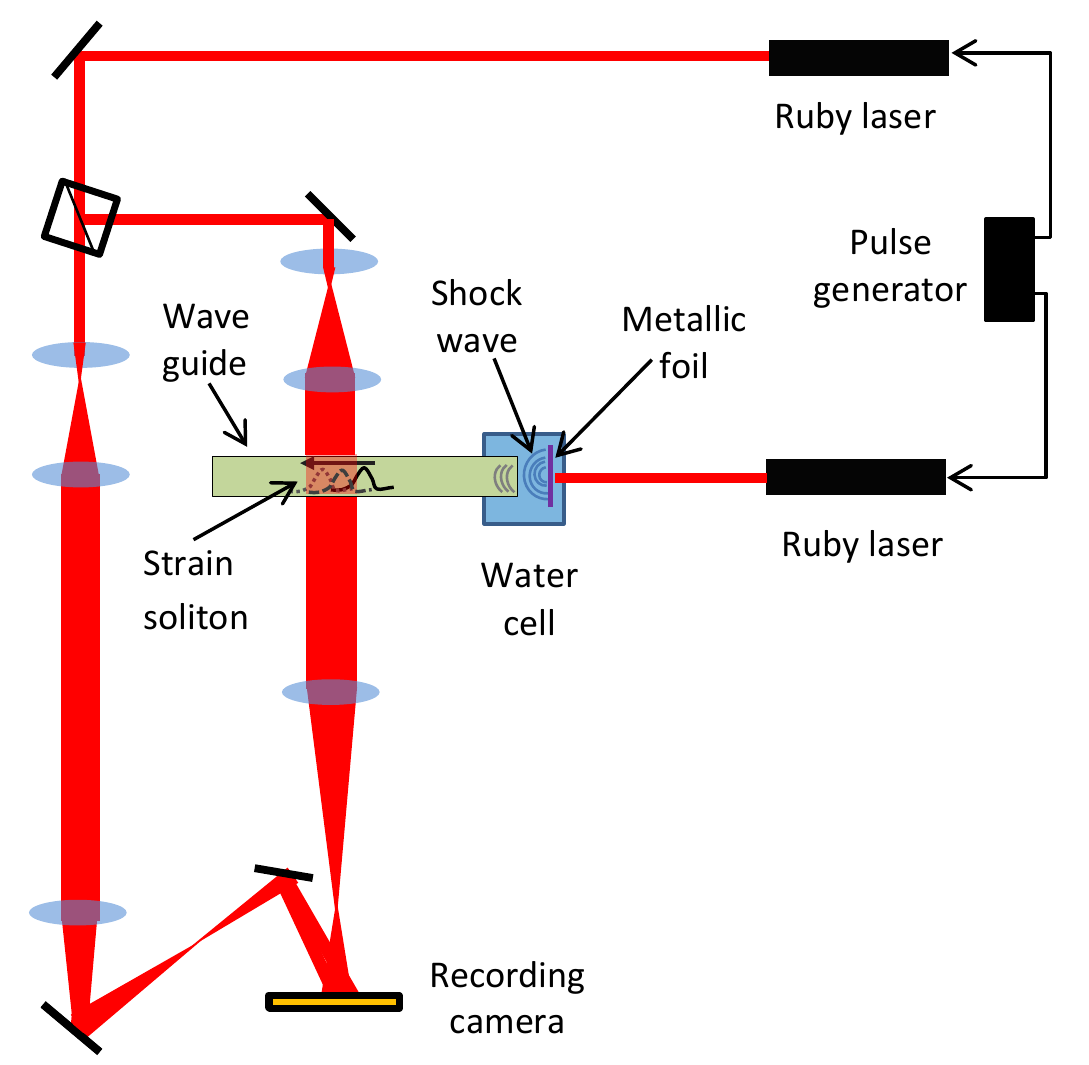}}
	\caption{Schematic of the experimental setup for soliton generation and detection.}
	\label{Excitation_setup}
\end{figure}

The soliton recording and monitoring of its evolution in the course of propagation along the waveguide is performed by means of either classical holographic interferometry or, most recently, by digital holography. These optical techniques operating in transmission configuration are completely noninvasive for a strain wave, they allow for recording of the entire wave pattern in one shot and provide most complete information on wave parameters.

In holographic interferometry the wave is characterized by the shift of carrier interference fringes, while in digital holography by the distribution of phase shift introduced by the strain wave to the recording wave front.  In both cases the maximal fringe shift or maximal phase shift provide data on soliton amplitude (the corresponding equations are derived in \cite{Strain2010,SPIE2018}), while fringe shift or phase shift distributions immediately give the soliton length. The soliton velocity is determined from measurements of its positions in the waveguide at different time moments set by the delay between pulses from the excitation and recording lasers. The detailed description of the experimental methodology can be found in the previous publications, see \cite{Dreiden-TP-2008,JAP2012,Strain2010,SPIE2018,APL2018}.  

The typical wave patterns illustrating the process of soliton generation in a polystyrene bar of square $1 \times 1 \ cm^2$ cross section are shown in Fig. ~\ref{soliton}. Note that experiments with rods and bars of comparable cross section have shown that soliton parameters and behaviour in a rod and in a bar do not differ much, but optical observations in a bar are much easier and clearer. Fig. ~\ref{soliton} (a) demonstrates an interferogram of the shock wave in water produced by laser evaporation of the metallic foil.  It was previously shown that this wave consists of a sharp compression peak, of about 0.2 $\mu m$ wide which is followed by a relatively long ($\sim 1\ mm$) rarefaction zone. When this wave enters the polystyrene bar it triggers the process of soliton formation, as shown in Fig. ~\ref{soliton} (b). At this early stage we can see a remainder of the initial shock wave which is followed by a long compression disturbance representing the transfer of the shock wave energy to the forming soliton. The wave pattern outside the bar presents accompanying waves in water. Our experiments showed that a soliton is formed at the distance of about $50 \ mm$ from the bar input. Fig. ~\ref{soliton} (c) demonstrates such a formed soliton at the distance of $70-120\ mm$ from the input. As can be seen from Fig. ~\ref{soliton} (c) the soliton in the polystyrene bar is a long trough-shaped compression wave, its main observable experimental parameters are the amplitude and width, which in this experiment were registered as $1.78 \times 10^{-4} $ and $34.2\ mm$, respectively, at the distance of $70 - 120\ mm$ from the input. The soliton speed was measured to be around $1800\ m/s$. At the moment, mathematical modelling of the soliton generation in this complicated fluid-structure  interaction experiment is an open problem.

\begin{figure}[htbp]
	\centering
	\fbox{\includegraphics[width=15 cm]{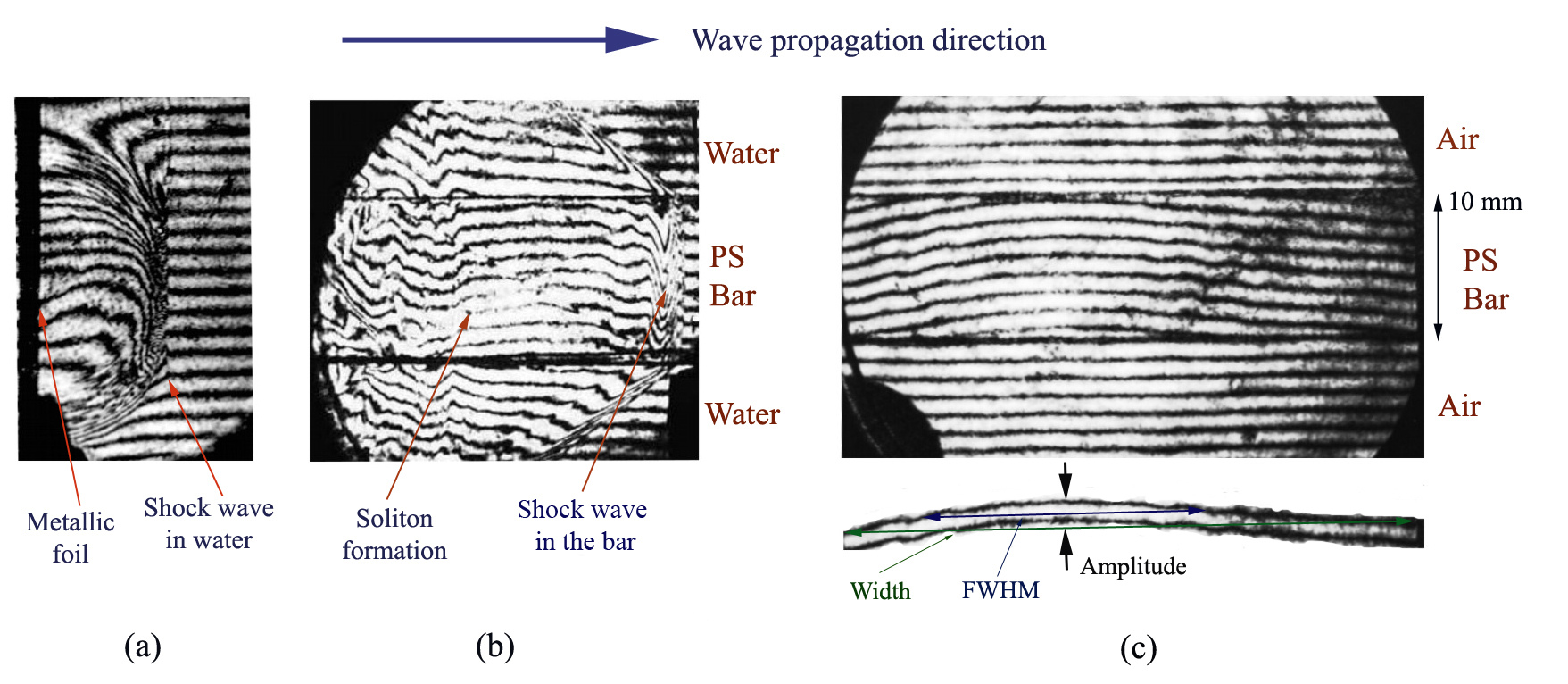}}
	\caption{Holographic interferograms of the shock wave in water (a), wave pattern in the bar at the distance of $0-40 \ mm$ from the input (b), bulk strain soliton at the distance of $70-120\  mm$ from the input (c).}
	\label{soliton}
\end{figure}



\section{Conclusions}

In this paper we derived, using a systematic asymptotic approach, two Boussinesq-type models describing long nonlinear longitudinal bulk strain waves in a rod of circular cross section with axially symmetric loading on the lateral boundary surface, and uni-axial (longitudinal) pre-stretch,  from the full nonlinear problem formulation within the scope of the Murnaghan model. The basic dynamical properties of the models, namely, linear dispersion curves and solitary waves solutions, have been analysed for the simplest case with no loading on the lateral surface, and no pre-stretch. The models have been compared between themselves, and with the existing model equations obtained, using some simplifications in the Lagrangian of the problem, by L.A. Ostrovsky and A.M. Sutin, and A.M. Samsonov and A.V. Porubov, respectively. We hope that Boussinesq-type equations derived in this paper will be useful for the modelling of the complicated generation process shown in Fig. ~\ref{soliton}, as well as other nonlinear dynamical processes in rods subjected to lateral loading and / or pre-stress.

\section{Acknowledgments}

We thank L.A. Ostrovsky and  A.V. Porubov for useful references and discussions. 
 F.E.G. and I.V.S. acknowledge the financial support from the Russian Science Foundation under the grant \# 17-72-20201. K.R.K. is grateful to the UK Institute of Mathematics and its Applications (IMA)  and the QJMAM Fund for Applied Mathematics for the financial support of her travel to the European Solid Mechanics Conference (ESMC2018) in Bologna, Italy in the summer of 2018  where parts of this work have been discussed and developed.

\section{Appendix A}

All Boussinesq-type equations discussed in the paper can be cast in the form
\begin{equation}
e_{tt} - c^2 e_{xx} = d_1 (e^2)_{xx} + d_2 e_{tttt} + d_3 e_{ttxx} + d_4 e_{xxxx},
\end{equation}
where $c$ and $d_i, i=\overline{1, 4}$ are some constants. Looking for the right- or left-propagating travelling-wave solutions
$$
e = e(\xi), \quad \mbox{where} \quad \xi = x \pm vt,
$$
we obtain the ordinary differential equation
\begin{equation}
(v^2 - c^2) e'' = d_1 (e^2)'' + (d_2 v^4 + d_3 v^2 + d_4) e^{IV}.
\label{ODE}
\end{equation}
Integrating this equation with respect to $\xi$ twice, and requiring that $e, e', e'', e''' \to 0$ as $\xi \to \pm \infty$, we obtain the equation
\begin{equation}
e'' = \frac{(v^2-c^2) e - d_1 e^2}{d_2 v^4 + d_3 v^2 + d_4},
\end{equation}
which can be viewed as Newton's equation of motion for a particle of unit mass in a potential field. The energy integral has the form
\begin{equation}
\frac 12 \left (e' \right )^2 -  \frac{3 (v^2-c^2) e^2 - 2 d_1 e^3}{6 (d_2 v^4 + d_3 v^2 + d_4)} = E,
\end{equation}
and the soliton solution corresponds to the zero energy level $E=0$. Separation of variables and the subsequent substitution
$$
e = \frac{3 (v^2 - c^2)}{2 d_1} {\rm sech}^2 \theta,
$$
where $\theta$ is a new variable,
allow one to obtain the solitary wave solution in the form
\begin{equation}
e = \frac{3 (v^2 - c^2)}{2 d_1} {\rm sech}^2 \left [\sqrt{\frac{v^2 - c^2}{4 (d_4 + d_3 v^2 + d_2 v^4)}} (x \pm v t)\right ]
\end{equation}
for the values of the parameter $v$ when this is a real-values function. The solution can be re-parametrised by the amplitude $A$:
\begin{equation}
e = A\  {\rm sech}^2 \left [\Lambda (x \pm v t)\right ],
\end{equation}
where
$$
\Lambda^2  = \frac{3 d_1 A}{2\left[ 9d_4 + 3d_3(3c^2 + 2Ad_1) + d_2(3c^2 + 2Ad_1)^2 \right]},  \quad v^2 = c^2 +  \frac 23 d_1 A.
$$


\end{document}